\documentclass[ aps,superscriptaddress,notitlepage]{revtex4-1}
\usepackage[utf8]{inputenc}
\usepackage[T1]{fontenc}
\usepackage{babel}
\usepackage{graphics}
\usepackage{amsfonts}
\usepackage{amssymb}
\usepackage{amsmath}
\usepackage{amsthm}
\usepackage{mathtools}
\usepackage{physics}
\usepackage{blindtext}
\usepackage{mathtools}
\usepackage[bookmarks=false,
  pdfpagelabels=false,
  hyperfootnotes=false,
  hyperindex=false,
  pageanchor=false,colorlinks=blue,]{hyperref}

\newcommand{\ba}{\begin{eqnarray}}
\newcommand{\ea}{\end{eqnarray}}

\begin{document}
\title{Thermodynamics of Reduced State of the Field}

\author{Stefano Cusumano}
\affiliation{International Centre for Theory of Quantum Technologies, University of Gdansk, 80-308 Gda{\'n}sk, Poland}
\email[Corresponding author: ]{stefano.cusumano@ug.edu.pl}

\author{{\L}ukasz Rudnicki}
\affiliation{International Centre for Theory of Quantum Technologies, University of Gdansk, 80-308 Gda{\'n}sk, Poland}
\affiliation{Center for Theoretical Physics, Polish Academy of Sciences, 02-668 Warszawa, Poland}





\begin{abstract}Recent years have seen the flourishing of research devoted to quantum effects on mesoscopic and macroscopic scales. In this context, in~\emph{Entropy} {\bf 2019}, \emph{21}, 705, a formalism aiming at describing macroscopic quantum fields, dubbed Reduced State of the Field (RSF), was envisaged. While, in the original work, a proper notion of entropy for macroscopic fields, together with their dynamical equations, was derived, here, we expand thermodynamic analysis of the RSF, discussing the notion of heat, solving dynamical equations in various regimes of interest, and showing the thermodynamic implications of these solutions.
\end{abstract}

\maketitle

\section{Introduction}

In recent years, a considerable attention has been given to the study of quantum phenomena on mesoscopic scale, as many physical systems that are nowadays fundamental for physical applications fall into this regime~\cite{datta,imamoglu}. The main characteristic of mesoscopic systems is that, while they are still large enough not to be considered purely quantum, they are neither small enough to ignore quantum effects.

Furthermore, while the behavior of macroscopic fields is well described by classical wave equations with coherent sources, incorporation of thermal and random sources into the field equations still represents an open problem~\cite{thorne}. As a matter of fact, the most common description of such a situation relies on the introduction of phenomenological terms, for example, terms describing the damping. This solution is not fully satisfactory from a theoretical point of view, as these extra terms do not give a correct thermodynamic description of such systems. On this basis, and on drawing from the fact that the ultimate description of any physical system should be given by quantum mechanics, the Reduced State of the Field (RSF) formalism was conceived~\cite{alicki}.

Since a completely quantum picture is generally too complex and, consequently, not convenient to treat macroscopic fields, the RSF aims at describing macroscopic waves using a coarse-grained version of the quantum formalism. Such a description allows one to retain the most important quantum features that would even emanate at macroscopic scale~\cite{alicki}, while, at the same time, mitigating the complexity that would have no effect beyond the microscopic realm. Interestingly, in the same spirit, one can answer the question being a sort of opposite to the former one, namely which features of the quantum evolution can be classified as classical~\cite{linowski}.

On the other hand, recent years have seen the flourishing of quantum thermodynamics~\cite{anders}, namely the study of thermodynamic phenomena on the quantum scale. This interest has been fostered by progressive miniaturization of electronic and optical devices, at the level where quantum phenomena cannot be ignored~\cite{pekola}.
We, therefore, observe a huge development of the field of quantum thermodynamics, where a wide range of topics is being covered, e.g., thermalization and heat transfer~\cite{podolski,cusumano,xu,ma}, quantum heat engines and refrigerators~\cite{linden, ono,kosloff,peterson,watanabe,yao,xiao}, and quantum batteries~\cite{binder,campaioli,andolina}.

The biggest advantage of the RSF formalism is that it provides both a suitable definition of entropy for radiation fields and dynamical equations describing the field which are in a closed form (do not depend on other degrees of freedom). It is, thus, of interest to see how thermodynamics intersects with the description of mesoscopic and macroscopic fields since, especially on the mesoscopic scale, one typically does not have full control over the system, yet quantum effects need to be taken into account in order to describe the system appropriately~\cite{rudnicki}.

In this paper, we want to explore how thermodynamic phenomena, such as heat exchange, fit the RSF formalism. Moreover, we want to analyze the behavior of the entropy of RSF \cite{alicki}, as its definition differs from the one usually found in the classical or quantum realms. The paper is organized as follows. In Section~\ref{sec:rsf_formalism}, we briefly review the RSF formalism, pointing out its main features. In Section~\ref{sec:definitions}, starting from the evolution equations of RSF, we consistently define the main thermodynamic quantities, such as internal energy, heat, and work.
Then, in Section~\ref{sec:thermodynamics}, we solve the equations of motion in some simple but relevant situations, highlighting the thermodynamic meaning of the different terms present therein. Finally, in Section~\ref{sec:conclusions}, we give our conclusions and some outlooks for future works.

\section{\label{sec:rsf_formalism}The RSF Formalism}

This section mostly follows Reference \cite{alicki}, since we summarize here the most important background and ingredients of the RSF formalism. In particular, all formulas appearing in this section are taken from Reference \cite{alicki}.

We start with classical electromagnetic field which, in a finite volume, is described by a set of modes $f_k(x;t)=e^{-i\omega_k t}f_k(x)$, where $x$ is the position, $k$ is a discrete index, and $\omega_k$ is the frequency at which the mode oscillates. In the first quantization picture, these modes represent eigenstates of the single-particle Hamiltonian of quasi-particles associated with the field. Under a proper normalization, these modes form an orthonormal basis of the single-particle Hilbert space, where the energy of each mode is equal to $\hbar\omega_k$.

In the second quantization picture, a pair of operators $\hat{a}_k,\hat{a}_k^\dag$ is associated to each mode $f_k$. Standard bosonic commutation relations hold:
\ba
\comm{\hat{a}_k}{\hat{a}_{k'}^\dag}=\delta_{kk'}\quad\comm{\hat{a}_k}{\hat{a}_{k'}}=\comm{\hat{a}_k^\dag}{\hat{a}_{k'}^\dag}=0,
\ea
so that the action of the annihilation and creation operators, on the vectors in the corresponding Fock space spanned by the orthonormal set $\{\ket{n}_k\}$, is
\ba
\hat{a}_k\ket{n_k}=\sqrt{n_k}\ket{n_k-1}\quad\hat{a}_k^\dag\ket{n_k}=\sqrt{n_k+1}\ket{n_k+1}.
\ea

The RSF formalism relies on a correspondence between operators acting on the single-particle Hilbert space and additive operators acting on the Fock space. The former can be written as:
\ba
\hat{b}=\sum_{k,k'}b_{kk'}\dyad{k}{k'},
\ea
where $\ket{k}\equiv\ket{f_k}$, while the corresponding additive observable in the Fock space is (We follow the 
 convention introduced in Reference \cite{alicki}, according to which operators in the Fock space are denoted by capital letters (with the density operator $\hat \rho$ being an exception), while operators acting on a single-particle Hilbert space are denoted by small letters.):
\ba
\hat{B}=\sum_{k,k'}b_{kk'}\hat{a}_k^\dag\hat{a}_{k'}.
\ea
Consequently, unitary operators $\hat{u}$ acting on the single-particle Hilbert space are in correspondence with multiplicative operators on the Fock space via:
\ba
\hat{u}=e^{i\hat{b}}\rightarrow\hat{U}=e^{i\hat{B}}.
\ea
From now on, we also use ``$\Tr$'' for trace operations in the Fock space and ``$\tr$'' for traces applied to the level of the RSF, i.e., on a single-particle Hilbert space.

The RSF description of the state of a macroscopic field is based on the couple $(\hat{r},\ket{\alpha})$, defined from the full quantum state of the field $\hat{\rho}$ in the Fock space as:
\begin{subequations}
\ba
&&\hat{r}=\sum_{k,k'}\Tr\left[\hat{\rho}\,\hat{a}^\dag_{k'}\hat{a}_k\right]\dyad{k}{k'}\coloneqq\sum_{k,k'}r_{kk'}\dyad{k}{k'},
\ea
\ba
&&\ket{\alpha}=\sum_k\Tr\left[\hat{\rho}\,\hat{a}_k\right]\ket{k}\coloneqq\sum_k\alpha_k\ket{k}.
\ea
\end{subequations}
The matrix $\hat{r}$ is a single-particle density operator, while the vector $\ket{\alpha}$ contains the information about the phase of the macroscopic field.

It is important to observe that the single-particle density operator is not normalized to unity but, rather, to the total number of particles in the state, i.e.,
\ba
\tr{\hat{r}}=N=\Tr{\hat{\rho}\hat{N}},\quad\hat{N}=\sum_k\hat{a}_k^\dag\hat{a}_k.
\ea
In fact, the same expectation-value identification holds for any additive observable
\ba
\tr{\hat{r}\hat{b}}=\Tr{\hat{\rho}\hat{B}}.
\ea

Furthermore, it turned out beneficial to define an another object, the {\it correlation matrix}
\ba
\hat{r}^\alpha=\hat{r}-\dyad{\alpha},\qquad\mathrm{where}\qquad\dyad{\alpha}=\sum_{k,k'}\alpha_k\alpha_{k'}^*\dyad{k}{k'},
\ea
which is a positive semi-definite operator being zero if and only if the state is coherent. Using this operator, it is then possible to give a suitable definition of entropy for macroscopic fields, which is
\ba
\label{eq:entropy_definition}
S[\hat{r}^\alpha]=k_B\tr\left[(\hat{r}^\alpha+1)\ln(\hat{r}^\alpha+1)-\hat{r}^\alpha\ln\hat{r}^\alpha\right].
\ea
This definition of entropy has an appealing feature of being always greater than or equal to zero, and being zero only when the RSF is coherent. This also highlights the fact that the coherent states are the only pure states in this formalism.

To shortly summarize the above, the RSF formalism is particularly suited to deal with situation where one does not have full quantum control of the system (we just control first and second moments, so to speak), as is in the case of macroscopic fields, but quantum effects are still visible. Having revised the RSF formalism and its main features, we are now ready to start thermodynamic considerations.

\section{\label{sec:definitions}Thermodynamics of the RSF}

In a usual scenario described by thermodynamics, one deals with a system $S$, often called the working fluid, interacting with one or more thermal baths, i.e., much larger systems with infinite heat capacity that are typically assumed to have a well-defined temperature. By changing the Hamiltonian, i.e., the energy, of the working fluid $S$ and letting it interact appropriately with the thermal baths, it is possible to extract work from the system (i.e., we have a heat engine) or to use work to transfer heat from a cold to a hot bath (i.e., we implement a refrigerator).

As in what follows, we will not be interested in a description of the thermal baths but, rather, in their action on the working fluid $S$. Therefore, we want to define heat and work only in terms of the state $S$, in the current context sufficiently well described by the couple $(\hat{r},\ket{\alpha})$.
In order to study the thermodynamics of a macroscopic field described under the RSF formalism, we first need to recall the dynamical equations describing the behavior of the field when it interacts with an external bath. This was already done in Reference~\cite{alicki}, where the system of equation for the RSF was derived from the standard expression for a map belonging to a so-called quasi-free dynamical semigroup~\cite{lendi,alicki2}, thus extending this concept to RSF formalism. The set of equations~\cite{alicki} describing the dynamics of the couple $(\hat{r},\ket{\alpha})$ can be derived from the equations describing the temporal evolution of the full state in Fock space $\hat{\rho}$ through:
\ba
\frac{d}{dt}r_{kk'}=\Tr{\hat{a}_{k'}^\dag\hat{a}_k\frac{d\hat{\rho}}{dt}},\qquad\frac{d}{dt}\alpha_k=\Tr{\hat{a}_k\frac{d\hat{\rho}}{dt}}.
\ea

 Considering a generic model of dynamics for $\hat \rho$, given by the evolution equation \cite{alicki}
\begin{align} \label{eq:Alicki_evolution_macroscopic}
\begin{split}
    \frac{d}{dt}\hat{\rho}=&
        -\frac{i}{\hbar}\big[\hat{H},\hat{\rho}\big]
        +\sum_{k=1}^{N}\big[\zeta_k\hat{a}_k^\dag-\zeta_k^*\hat{a}_k,\hat{\rho}\big] +\sum_{k,k'=1}^{N}\Gamma^{k'k}_\downarrow\left(
        \hat{a}_{k}\hat{\rho}\,\hat{a}_{k'}^\dag
        -\frac{1}{2}\big\{\hat{a}_{k'}^\dag\hat{a}_{k},\hat{\rho}\big\}\right)\\
    &+\sum_{k,k'=1}^{N}\Gamma^{k'k}_\uparrow\left(
        \hat{a}_{k'}^\dag\hat{\rho}\,\hat{a}_{k}
        -\frac{1}{2}\big\{\hat{a}_{k}\hat{a}_{k'}^\dag,\hat{\rho}\big\}\right)
    +\int\mu(du)\left(\hat{U}\hat{\rho}\,\hat{U}^\dag-\hat{\rho}\right),
\end{split}
\end{align}
which includes the presence of a coherent source, a thermal bath, and random scattering, one can write the following equations for the couple $(\hat{r},\ket{\alpha})$ (Note that
 the anticommutator terms, in comparison with Reference \cite{alicki}, have been divided by $2$. See Reference \cite{linowski} for details.):
\begin{subequations}
\begin{align}
\label{eq:r_dynamics}
\frac{d}{dt}\hat{r}&=-\frac{i}{\hbar}[\hat{h},\hat{r}]+(\dyad{\zeta}{\alpha}+\dyad{\alpha}{\zeta})+\frac{1}{2}\left\{(\hat{\gamma}_{\uparrow}-\hat{\gamma}_{\downarrow}),\hat{r}\right\}+\hat{\gamma}_{\uparrow}\nonumber\\&+\int \mu(du)(\hat{u}\hat{r}\hat{u}^\dag-\hat{r}),
\end{align}
\ba
\label{eq:alpha_dynamics}
 \frac{d}{dt}\ket{\alpha}=-\frac{i}{\hbar}\hat{h}\ket{\alpha}+\ket{\zeta}+\frac{1}{2}(\hat{\gamma}_{\uparrow}-\hat{\gamma}_{\downarrow})\ket{\alpha}+\int\mu(du)(\hat{u}-1)\ket{\alpha}.
\ea
\end{subequations}
Let us start by explaining the meaning of each term in (\ref{eq:Alicki_evolution_macroscopic}) viz. Equations~(\ref{eq:r_dynamics}) and (\ref{eq:alpha_dynamics}).
In the dynamical equation for $\hat{r}$, we first find the commutator of $\hat{r}$ with the single-particle Hamiltonian $\hat{h}=\hbar\sum_k\omega_k\dyad{k}$ stemming from $\hat{H}= \hbar\sum_k \omega_k \hat{a}_k^\dag\hat{a}_k$, and this term describes nothing but the standard unitary dynamics induced by the free Hamiltonian. Next, we find the term $\dyad{\zeta}{\alpha}$, which describes the effect of a coherent source, and, thus, also depends on the phase of the system $\ket{\alpha}$. Then, we can see the anticommutator term with the operators 
\ba
\hat{\gamma}_{\updownarrow}=\sum_{k,k'}\Gamma^{kk'}_{\updownarrow}\dyad{k}{k'},
\ea
describing stimulated absorption and emission processes, while the isolated term $\hat{\gamma}_{\uparrow}$ describes spontaneous emission processes. The coefficients $\Gamma_{\updownarrow}^{kk'}$ encode the information about the state of the thermal bath and its interaction with the system. Finally, the integral term describes the effect of random scattering phenomena, where the operators $\hat{u}$ are unitary. Similar considerations apply to the dynamical equation for $\ket{\alpha}$. Note also that, although the usual single particle approach is one 
where recursive systems of equations are truncated through appropriate approximations or boundary conditions, in the RSF approach, one deals with a closed system of equation, a feature that greatly simplifies the study of the dynamics of a macroscopic field.

As the entropy is defined in terms of the correlation matrix $\hat{r}^{(\alpha)}$, it is also useful to derive the dynamical equation for this quantity. Since $\hat{r}^{(\alpha)}=\hat{r}-\dyad{\alpha}$, we only need to compute the time derivative of $\dyad{\alpha}$ using Equation~\eqref{eq:alpha_dynamics}:
\begin{align}
\nonumber
\frac{d}{dt}\dyad{\alpha}=&\left(\frac{d}{dt}\ket{\alpha}\right)\bra{\alpha}+\ket{\alpha}\left(\frac{d}{dt}\bra{\alpha}\right)\\
\nonumber
=&-\frac{i}{\hbar}\comm{\hat{h}}{\dyad{\alpha}}+(\dyad{\alpha}{\zeta}+\dyad{\zeta}{\alpha})+\frac{1}{2}\acomm{(\hat{\gamma}_\uparrow-\hat{\gamma}_\downarrow)}{\dyad{\alpha}}\\
+&\int\mu(du)(\hat{u}\dyad{\alpha}+\dyad{\alpha}\hat{u}^\dag)-2\dyad{\alpha}),
\end{align}
from which we can write the dynamical evolution for the correlation matrix $\hat{r}^{(\alpha)}$ as
\ba
\nonumber
&&\frac{d}{dt}\hat{r}^{(\alpha)}=-\frac{i}{\hbar}[\hat{h},\hat{r}^{(\alpha)}]+\frac{1}{2}\left\{(\hat{\gamma}_{\uparrow}-\hat{\gamma}_{\downarrow}),\hat{r}^{(\alpha)}\right\}+\hat{\gamma}_{\uparrow}\\
\label{eq:corr_dynamics}
&&\qquad\quad+\int \mu(du)\left(\hat{u}\hat{r}^{(\alpha)}\hat{u}^\dag-\hat{r}^{(\alpha)}\right)+\int \mu(du)(\hat{u}-1)\dyad{\alpha}(\hat{u}^\dagger-1).
\ea
From this equation, we can see that the dynamics of the correlation matrix are not influenced by the presence of coherent sources. Consequently, the entropy $S[\hat{r}^{(\alpha)}]$ is also invariant with respect to coherent evolution. This feature of the theory is associated with the fact that we are dealing with a mesoscopic or macroscopic system, where, in fact, we do not have access to all degrees of freedom~\cite{alicki}. In particular, the single-particle Hamiltonian $\hat h$ does not carry the whole content of the Hamiltonian in the Fock space which also contains contributions due to the displacement. In view of this, we define the internal energy as
\ba
\label{eq:internal_energy_definition}
U=\tr[\hat{h}\hat{r}^{(\alpha)}]\equiv\tr[\hat{h}\hat{r}]-\bra{\alpha}\hat{h}\ket{\alpha}.
\ea
This definition is motivated by the form of the entropy in Equation~\eqref{eq:entropy_definition} and from the related discussion in Reference~\cite{alicki}: as the definition of entropy relies on the effective degree of control that one has over the physical system under examination, the same should apply to other quantities of interest.
Since, in the RSF formalism, the entropy is invariant under the application of the Weyl displacement operator, one could expect the internal energy to follow the same behavior. In particular, if, for instance, we were to define the internal energy in the ``intuitive'' way as $\tr[\hat{h}\hat{r}]$, then displacement would be a process implying heat absorption from the system, with no change of entropy. In Section \ref{sec:thermodynamics}, we are going to show that this issue is resolved by Equation~\eqref{eq:internal_energy_definition}, and that, thanks to this definition, we are able to define properly the free energy of the system. Last but not least, let us emphasize that the internal energy of the system is a notion which depends on an arbitrary choice in 
which degrees of freedom describe the system and which belong to its environment.

Using the notion of internal energy in Equation~\eqref{eq:internal_energy_definition}, one has a natural decomposition
\ba
dU=\tr[\frac{d\hat{h}}{dt}\hat{r}^{(\alpha)}]dt+\tr[\hat{h}\frac{d\hat{r}^{(\alpha)}}{dt}]dt=dW+\delta Q.
\ea
Two observations are in place here. First of all, the single particle Hamiltonian is time independent by construction. This is because the frequencies, as well as the eigenmode basis, of the Hamiltonian, are not under control and do not vary over time due to the dynamics of the sole field. Therefore, for generic macroscopic fields, there is no work, just the heat. Work would require an engineered variant of time evolution, i.e., one can perform (extract) work on (from) the system only by changing the frequencies $\omega_k$.

Second of all, only the scattering term couples $\hat{r}^{(\alpha)}$ with $\ket{\alpha}$ in Equation \ref{eq:corr_dynamics}. This feature in a salient way distinguishes the scattering processes from the other processes subsumed in the dynamical equations. Within a thermodynamic description, which is solely based here on the correlation matrix, the scattering belongs to a different (more complex) class of (likely non-equilibrium) processes. The latter property, however, would strongly depend on the measure $\mu(du)$ chosen. Perhaps, for the invariant Haar measure, the situation would simplify, still, the aforementioned coupling will be there. 

Therefore, we believe that the scattering processes deserve a separate and detailed treatment. Consequently, here, we shall neglect random scattering terms, with the goal of delineating the heat exchange and entropy production due to other processes. Under this simplifying assumption, the heat exchanged is equal to
\ba
\nonumber
&&\delta Q=\tr[\hat{h}\frac{d\hat{r}^{(\alpha)}}{dt}]dt=\frac{1}{2}\tr[\hat{r}^{(\alpha)}\acomm{\hat{\gamma}_\uparrow-\hat{\gamma}_\downarrow}{\hat{h}}]dt+\tr[\hat{h}\hat{\gamma}_{\uparrow}]dt\\
\label{eq:corr_heat}
&&\qquad=\hbar\sum_{k,k'}\frac{\omega_k+\omega_{k'}}{2}r_{kk'}^{(\alpha)}\left(\Gamma_\uparrow^{k'k}-\Gamma_\downarrow^{k'k}\right)dt+\hbar\omega_k\Gamma_\uparrow^{kk}dt,
\ea
that is, it only depends on interactions with the thermal bath. In particular, the second term on the right-hand side of Equation~\eqref{eq:corr_heat} is responsible for the equilibration process towards the equilibrium populations dictated by the bath structure, while the first term describes heat exchanges due to changes in the modes' occupations happening because of the interaction with the bath.

The variation of the entropy in time is also found to be
\ba\label{entder}
\frac{d}{dt}S[\hat{r}^{(\alpha)}]=k_B \tr[\frac{d\hat{r}^{(\alpha)}}{dt}\ln(\frac{\hat{r}^{(\alpha)}+1}{\hat{r}^{(\alpha)}})].
\ea
We use the notation in which the fraction of non-negative operators needs to be understood in terms of their eigenvalues. This is possible because, whenever some eigenvalue approaches $0$, the time derivative also vanishes, killing the potential singularities \cite{derivative}.

Note that the trace of $\hat{r}^{(\alpha)}$ does not need to be constant in time.
 For a quasi-static process, in which the state $\hat \rho$ is always in thermal equilibrium, the correlation matrix is always of the form
\ba
\label{eq:thermal_corr}
\hat{r}^{(\alpha)}=\frac{1}{e^{\beta\hat{h}}-1}.
\ea
Since, in this case,
\ba
\label{eq:thermal_corr}
\ln(\frac{\hat{r}^{(\alpha)}+1}{\hat{r}^{(\alpha)}})=\beta\hat{h},
\ea
we recover the equality from standard thermodynamics
\ba
dS=k_B \beta\delta Q.
\ea
This observation further strengthens our definition of work and heat. Moreover, for a non-quasi-static process, one has that $\hat{r}^{(\alpha)}$ is not of the form in Equation~\eqref{eq:thermal_corr}; thus, one has also entropy production.
\section{\label{sec:thermodynamics}Some Examples of RSF Thermodynamics}

In the following subsections, we want to solve the dynamical Equations~(\ref{eq:r_dynamics}) and \mbox{(\ref{eq:alpha_dynamics})} under various circumstances where some of the terms are absent or can be simplified, thus highlighting their thermodynamic meaning.

\subsection{Free Dynamics of the RSF}
The simplest, and almost trivial, case that one can analyze is the one where no interaction with either a coherent source or a thermal bath is present, so that the dynamics of the RSF is fully described by the Hamiltonian term alone.
Assuming the Hamiltonian in the Fock space to be:
\ba
\hat{H}=\sum_k\hbar\omega_k\hat{a}_k^\dag\hat{a}_k\rightarrow\hat{h}=\sum_k\hbar\omega_k\dyad{k},
\ea
we can explicitly write down the equations governing the matrix elements $r_{kk'}(t)$ and the vector components $\alpha_k(t)$ as:
\ba
\frac{d}{dt}r_{kk'}(t)=-i(\omega_k-\omega_{k'})r_{kk'}(t),\quad\frac{d}{dt}\alpha_k(t)=-i\omega_k\alpha_k(t).
\ea
The solutions to these equations are easily found:
\ba
r_{kk}(t)=r_{kk}(0);\quad r_{kk'}=e^{-i(\omega_k-\omega_{k'})t}r_{kk'}(0);\quad\alpha_k(t)=e^{-i\omega_kt}\alpha_k(0).
\ea

These solutions imply that, under purely free dynamics, the populations stay constant, while the coherences among them rotate at a frequency equal to the detuning between the modes. Finally, the components of the phase vector $\ket{\alpha}$ rotate at the corresponding frequency.
In accordance with Equation~\eqref{eq:corr_heat}, there is no heat exchange, as there is no thermal bath. An important fact to be noted is that, as in Equation~\eqref{eq:corr_dynamics}, the correlation matrix depends on the Hamiltonian $\hat{h}$ only through the commutator term, and the entropy is unchanged under purely Hamiltonian dynamics, since the eigenvalues of $\hat{r}^{(\alpha)}$ are \mbox{left unchanged.}



\subsection{RSF Dynamics in Presence of a Coherent Source}

We now want to solve Equations~(\ref{eq:r_dynamics}) and (\ref{eq:alpha_dynamics}) subject to a coherent source, but still without a thermal bath, so that we get:
\ba
&&\frac{d}{dt}\hat{r}=-\frac{i}{\hbar}[\hat{h},\hat{r}]+(\dyad{\alpha}{\zeta}+\dyad{\zeta}{\alpha}),\\
&&\frac{d}{dt}\ket{\alpha}=-\frac{i}{\hbar}\hat{h}\ket{\alpha}+\ket{\zeta},
\ea
where $\ket{\zeta}=\sum_k\zeta_k\ket{k}$. We can easily get the dynamical equations for the matrix elements:
\ba
\label{eq:coherent_r}
&&\frac{d}{dt}r_{kk'}=-i(\omega_k-\omega_{k'})r_{kk'}+\left(\alpha_k\zeta_{k'}^*+\alpha_{k'}^*\zeta_k\right),\\
\label{eq:coherent_alpha}
&&\frac{d}{dt}\alpha_k=-i\omega_k\alpha_k+\zeta_k.
\ea
Solving the second equation first, we get
\ba
\alpha_k(t)=e^{-i\omega_kt}\alpha_k(0)-i\frac{\zeta_k}{\omega_k}(1-e^{-i\omega_kt}),
\ea
so that the $\hat{r}$ matrix elements are
\ba
r_{kk'}(t)=e^{-i(\omega_k-\omega_{k'})t}\left[r_{kk'}(0)+\int_0^tds\,e^{i(\omega_k-\omega_{k'})s}\big(\alpha_k(s)\zeta_{k'}^*+\alpha_{k'}^*(s)\zeta_k\big)\right].
\ea
After we perform the integral, we get
\ba
\nonumber
r_{kk'}(t)&=&e^{-i(\omega_k-\omega_{k'})t}\left(r_{kk'}(0)+\frac{\zeta_k\zeta_{k'}}{\omega_k\omega_{k'}}\right)+\frac{\zeta_k\zeta_{k'}^*}{\omega_k\omega_{k'}}\left(1- e^{-i\omega_k t}- e^{i\omega_{k'}t}\right)\\
&+&i\left(\frac{\alpha_k(0)\zeta_{k'}^*}{\omega_{k'}}e^{-i\omega_kt}-\frac{\alpha_{k'}^*(0)\zeta_k}{\omega_k}e^{i\omega_{k'}t}\right).
\ea

Let us consider the case where the initial phase vector $\ket{\alpha}$ is null, i.e., $\alpha_k(0)=0$, for all $k$.
In this case, the solution for the phase and the matrix elements of $\hat{r}$ reads:
\ba
&&\alpha_k(t)=-i\frac{\zeta_k}{\omega_k}(1-e^{-i\omega_k t}),
\ea
\ba
r_{kk'}(t)=e^{-i(\omega_k-\omega_{k'})t}\left(r_{kk'}(0)+\frac{\zeta_k\zeta_{k'}^*}{\omega_k\omega_{k'}}\right)+\frac{\zeta_k\zeta_{k'}^*}{\omega_k\omega_{k'}}\left(1- e^{-i\omega_k t}- e^{i\omega_{k'}t}\right).
\ea
The latter result, for the diagonal elements $r_{kk}(t)$, reduces to
\ba
r_{kk}(t)=r_{kk}(0)+2\frac{|\zeta_k|^2}{\omega_k^2}\left(1-\cos{\omega_kt}\right).
\ea
This implies that the populations oscillate around the average values $r_{kk}(0)+|\zeta_k|^2/\omega_k^2$. Of course, the correlation matrix remains constant (also if initial $\ket{\alpha}$ is not null), so does \mbox{the entropy.}



\subsection{Dynamics of the RSF in Presence of a Coherent Source and a Thermal Bath}

Let us now consider the case where also a dissipation term is present, i.e., we want to analyze the case where the system interacts with both a coherent source and a heat bath.
In this case, the dynamical equations for $\hat{r}$ and $\ket{\alpha}$ are:
\ba
&&\frac{d\hat{r}}{dt}=-\frac{i}{\hbar}[\hat{h},\hat{r}]+(\dyad{\alpha}{\zeta}+\dyad{\zeta}{\alpha})+\frac{1}{2}\{(\hat{\gamma}_{\uparrow}-\hat{\gamma}_{\downarrow}),\hat{r}\}+\hat{\gamma}_{\uparrow},\\
&&\frac{d}{dt} \ket{\alpha}=-\frac{i}{\hbar}\hat{h}\ket{\alpha}+\ket{\zeta}+\frac{1}{2}(\hat{\gamma}_{\uparrow}-\hat{\gamma}_{\downarrow})\ket{\alpha},
\ea
where the operators $\hat{\gamma}_{\updownarrow}$ have already been defined as
\ba
\hat{\gamma}_{\updownarrow}=\sum_{k,k'}\Gamma^{kk'}_{\updownarrow}\dyad{k}{k'}.
\ea
Let us remind that the matrix elements $\Gamma_{\updownarrow}^{kk'}$ are the particle creation and decay rates that can be derived using the Fermi golden rule. Under the typical Born, Markov, and secular approximations, the operators $\hat{\gamma}_{\updownarrow}$ become diagonal
\ba
\hat{\gamma}_{\updownarrow}=\sum_k\Gamma_{\updownarrow}^{k}\dyad{k},
\ea
where the rates $\Gamma_{\updownarrow}^k$, due to the thermal character of the bath, are related via
\ba
\frac{\Gamma_{\uparrow}^k}{\Gamma_{\downarrow}^k}=e^{-\frac{\hbar\omega_k}{k_BT}},
\ea
with $k_B$ being the Boltzmann constant, and $T$ being the temperature of the heat bath.

In this case, the dynamical equations for the RSF become:
\ba
&&\frac{dr_{kk'}}{dt}=-i(\omega_\kappa-\omega_{k'})r_{kk'}-\frac{1}{2}\left(\frac{\Gamma_\downarrow^k}{Z_k}+\frac{\Gamma_\downarrow^{k'}}{Z_{k'}}\right)r_{kk'}+\delta_{kk'}\Gamma_{\uparrow}^k+\left(\alpha_k\zeta_{k'}^*+\alpha_{k'}^*\zeta_k\right),\\
&&\hspace{100pt}\frac{d\alpha_k}{dt}=-i\omega_k\alpha_k-\frac{\Gamma_\downarrow^k}{2Z_k}\alpha_k+\zeta_k,
\ea
where we have defined $Z_k=\left(1-e^{-\beta\hbar\omega_k}\right)^{-1}$. These equations are of the same form as Equations (\ref{eq:coherent_r}) and (\ref{eq:coherent_alpha}). This can be noted by defining complex frequencies $\tilde{\omega}_k=\omega_k-i\Gamma_\downarrow^k/2Z_k$. In this notation, we get:
\ba
&&\frac{dr_{kk'}}{dt}=-i(\tilde{\omega}_k-\tilde{\omega}_{k'}^*)r_{kk'}+\delta_{kk'}\Gamma_\uparrow^k+(\alpha_k\zeta_{k'}^*+\alpha_{k'}^*\zeta_k),\\
&&\hspace{60pt}\frac{d\alpha_k}{dt}=-i\tilde{\omega}_{k}\alpha_k+\zeta_k,
\ea
so that one can immediately write down the solution to the second equation as:
\ba
\alpha_k(t)=e^{-i\tilde{\omega}_kt}\alpha_k(0)-i\frac{\zeta_k}{\tilde{\omega}_k}\left(1-e^{-i\tilde{\omega}_kt}\right),
\ea
which implies that the phases $\alpha_k$ are driven towards their steady-state values
\ba
\label{eq:steady_alpha}
\alpha_{k}^{\textnormal{steady}}=-i\frac{\zeta_k}{\tilde{\omega}_k}.
\ea
As for the matrix elements $r_{kk'}$, one finds:
\ba
\nonumber
r_{kk'}(t)=e^{-i(\tilde{\omega}_k-\tilde{\omega}_{k'}^*)t}\left[r_{kk'}(0)+\int_0^tds\,e^{i(\tilde{\omega}_k-\tilde{\omega}_{k'}^*)s}\left(\alpha_k(s)\zeta_{k'}^*+\alpha_{k'}^*(s)\zeta_k+\delta_{kk'}\Gamma_\uparrow^k\right)\right],
\ea
and, consequently,
\ba
\nonumber
r_{kk'}(t)&=&e^{-i(\tilde{\omega}_k-\tilde{\omega}_{k'}^*)t}\left(r_{kk'}(0)+\frac{\zeta_k\zeta_{k'}}{\tilde{\omega}_k\tilde{\omega}_{k'}^*}\right)+\frac{\zeta_k\zeta_{k'}^*}{\tilde{\omega}_k\tilde{\omega}_{k'}^*}\left(1- e^{-i\tilde{\omega}_k t}- e^{i\tilde{\omega}_{k'}^*t}\right)\\
&+&i\left(\frac{\alpha_k(0)\zeta_{k'}^*}{\tilde{\omega}_{k'}^*}e^{-i\tilde{\omega}_kt}-\frac{\alpha_{k'}^*(0)\zeta_k}{\tilde{\omega}_k}e^{i\tilde{\omega}_{k'}^*t}\right)+\delta_{kk'}e^{-\beta\hbar\omega_k}Z_k\left(1-e^{-\frac{\Gamma_\downarrow^k}{Z_k}t}\right).
\ea

It is of particular interest to see the steady values of the matrix elements $r_{kk'}$,
\ba
r_{kk'}^{\textnormal{steady}}=\frac{\zeta_k\zeta_{k'}^*}{\tilde{\omega}_k\tilde{\omega}_{k'}^*}+\delta_{kk'}e^{-\beta\hbar\omega_k}Z_k.
\ea
From this steady-state solution, together with Equation~\eqref{eq:steady_alpha}, we can compute the associated correlation matrix $\hat{r}^{(\alpha)}$, for which one simply obtains:
\ba
\label{eq:alpha_steady}
r_{kk'}^{(\alpha)\textnormal{steady}}=r_{kk'}^{\textnormal{steady}}-\dyad{\alpha^{\textnormal{steady}}}_{kk'}=\delta_{kk'}e^{-\beta\hbar\omega_k}Z_k=\frac{1}{e^{\beta\hat{h}}-1}.
\ea
From this result, one can see clearly what was already noted in Reference~\cite{alicki}, namely that, in presence of random scattering (which is absent in this case) or a thermal environment with temperature different from zero, it is impossible to obtain a coherent state, and that only an initial pure state remains pure when the above conditions are met.

\begin{figure}[h]
\includegraphics[scale=0.6]{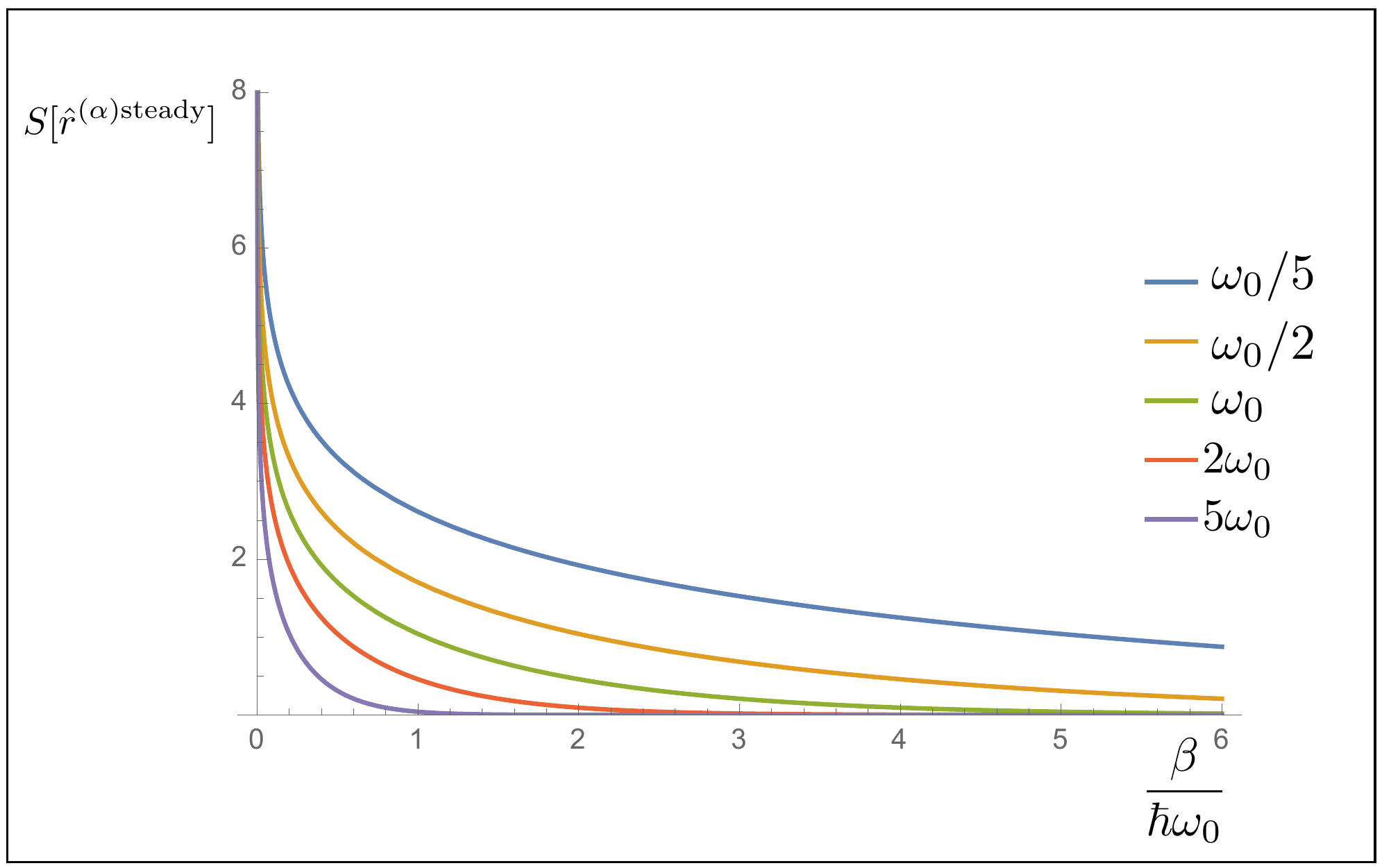}
\caption{In this plot, the entropy as a function of temperature $\beta$ is shown. The various lines are plotted using different frequency, so that one can see that the low frequency modes contribute more to the entropy, especially at low temperatures.}
\label{fig:entropy_beta}
\end{figure}

Next, we express the entropy of the steady state as a function of $\beta$ (we set $k_B=1$):
\ba
\nonumber
S[\hat{r}^{(\alpha)\textnormal{steady}}](\beta)&=&\tr[(\hat{r}^{(\alpha)\textnormal{steady}}+1)\ln(\hat{r}^{(\alpha)\textnormal{steady}}+1)-\hat{r}^{(\alpha)\textnormal{steady}}\ln\hat{r}^{(\alpha)\textnormal{steady}}]\\
\nonumber
&=&\tr[\beta\hat{h}\hat{r}^{(\alpha)\textnormal{steady}}]+\tr[\ln(\hat{r}^{(\alpha)\textnormal{steady}}+1)]\\
\label{eq:steady_entropy}
&=&\beta U+\tr[\ln(\hat{r}^{(\alpha)\textnormal{steady}}+1)],
\ea
as it can be found using Equation~\eqref{eq:alpha_steady} and going through some algebra.
One can immediately see that the entropy depends on the temperature, both through the partition functions and the occupation numbers of the modes. We plot in Figure~\ref{fig:entropy_beta} the entropy as a function of the temperature $\beta$ for different values of the frequency. From the plot, it can be seen that lower frequency modes have a greater entropy than the modes with higher frequency.

The Equation~\eqref{eq:steady_entropy} can also be rearranged as:
\ba
U-\beta^{-1}S=-\frac{1}{\beta}\tr[\ln(\hat{r}^{(\alpha)\textnormal{steady}}+1)],
\ea
so that, in this way, we are driven to define the equilibrium-free energy $F_{eq}$
\ba
F_{eq}=-\frac{1}{\beta}\tr[\ln(\hat{r}^{(\alpha)\textnormal{steady}}+1)]=-\frac{1}{\beta}\sum_{k}\ln Z_k.
\ea
 This is exactly the sum of the equilibrium-free energies of each mode.
 We can then define the free energy as:
 \ba
 \nonumber
 F&=&U-\beta^{-1}S=\tr[\hat{r}^{(\alpha)}\hat{h}]-\frac{1}{\beta}\tr[(\hat{r}^{(\alpha}+1)\ln(\hat{r}^{(\alpha}+1)-\hat{r}^{(\alpha)}\ln\hat{r}^{(\alpha)}]\\
 &=&\tr[\hat{r}^{(\alpha)}\left(\hat{h}-\frac{1}{\beta}\ln\left(\frac{\hat{r}^{(\alpha)}+1}{\hat{r}^{\alpha}}\right)\right)]-\frac{1}{\beta}\tr[\ln(\hat{r}^{(\alpha)}+1)]\\
 &=&F_{neq}+F_{eq},
 \ea
where we have introduced the non-equilibrium-free energy
\ba
F_{neq}=\tr[\hat{r}^{(\alpha)}\left(\hat{h}-\frac{1}{\beta}\ln\left(\frac{\hat{r}^{(\alpha)}+1}{\hat{r}^{\alpha}}\right)\right)].
\ea
Thus, we see how, in the presence of a thermal bath, and using the definition of internal energy of Equation~\eqref{eq:internal_energy_definition}, we are able to define in a reasonable way the free energy, both ``in and out'' of equilibrium. Clearly, the proposed notion of free energy is somehow attached to the specific case of macroscopic fields. This is to be expected since, in the RSF formalism, one assumes the lack of control over certain (actually, many) degrees of freedom. Therefore, in its spirit, our approach does not differ from descriptions of other physical situations, such as the modeling of magnetic~\cite{zhang1,agarwal} and molecular~\cite{zhang2,zhang3} systems, where adjustments are necessary in order to account for the specific properties of the system under examination.

\section{\label{sec:conclusions}Conclusions}

In this paper, we explored how to define thermodynamic quantities in the RSF formalism, given its definition of entropy. We also showed some examples of dynamical regimes that allowed us to explicitly compute the quantities of our interest, such as energy, heat, work, and other thermodynamic functionals.

Starting from the definition of entropy given in Reference~\cite{alicki}, we gave a reasonable definition of internal energy, heat, and work. We were able to show that, in a quasi-static equilibrium process, our definition of heat gave the proper increase of entropy, and then we defined the equilibrium and non-equilibrium-free energy.

It would be interesting in the future to further explore how to describe other thermodynamic phenomena under this formalism, such as work extraction from heat engines and work storage in batteries. This would surely help to further clarify how thermodynamics should be described at mesoscopic scales, as well as to individuate possible issues to be solved in this regime. Last but not least, scattering terms deserve a careful, separate consideration.

 \acknowledgments{We acknowledge support by the Foundation for Polish Science (IRAP project, ICTQT, Contract No. 2018/MAB/5, co-financed by the EU within the Smart Growth Operational Program).}


\end{document}